# On the Possibility of Superfast Charge Transfer in DNA

## Lakhno V. D., Sultanov V. B.

*Institute of Mathematical Problems of Biology, RAS, 142290, Pushchino, Institutskaya str., 4, Russia*

*Abstract.* Numerous experiments on charge transfer in DNA yield a contradictory picture of the transfer: on the one hand they suggest that it is a very slow process and the charge is almost completely localized on one Watson-Crick pair, but on the other hand they demonstrate that the charge can travel a very large distance. To explain this contradiction we propose that superfast charge transitions are possible between base pairs on individual DNA fragments resulting in the establishment of a quasi-equilibrium charge distribution during the time less than that of charge solvation. In other words, we hypothesize these states irrespective of the nature of a mechanism responsible for their establishment, whether it be a hopping mechanism, or a band mechanism, or superexchange, or polaron transport, etc., leaving aside the debates of which one is more advantageous. We discuss qualitative differences between the charge transfer in a dry DNA and that in a solution. In a solution, of great importance is the charge solvation which decreases the transfer rate $10^7 \div 10^8$ times as compared with a dry DNA. We consider the conditions under which the superfast charge transfer in a DNA leading to quasi-equilibrium distributions of polarons in a duplex is possible. Comparison of calculated quasi-equilibrium distributions with the experiment testifies to the possibility of superfast tunnel transitions of a hole in a DNA duplex in a solution.

*Key words:* Holstein Hamiltonian, hole, solvation.

The possibility of a long-range charge transfer in a DNA has been a subject of discussion for as many as 15 years. Presently this possibility is practically assured. It is proved by numerous experiments on charge transfer in DNA carried out during the last 1.5 decades [1]. However, the mechanism of such transfer is still unclear. Meanwhile it must be clarified not only in view of the general-biological importance of the problem which is associated with damage, mutation and repair processes in DNA [2–5], but also due to the fact that presently DNA is considered as a basis for construction of circuitry elements in nanobiological devices [6, 7]. In the absence of a solvent, in dry conditions relevant to most of proposed nanobioelectronic setups, possible mechanisms of charge transfer include: polaron or soliton transport [8–10], variable range hopping [11, 12], bandlike electronic or hole transport [13, 14], combined hopping superexchange mechanisms [15].

The overwhelming majority of experiments on charge transfer in DNA are carried out in a solution when the contribution of the solvent is important. In recent papers [16, 17] it has been shown that the effect of solvation leads to a strong localization of a charge on an individual nucleotide pair which rules out the band mechanism of the hole conductivity even in homogeneous nucleotide sequences. The polaron mechanism of transfer also becomes problematic since for small-radius polarons practically entirely localized in a deep potential well on one nucleotide, the probability of temperature jumps becomes very small. Only due to a very small rate of a hole-water reaction (and subsequent site-selective strand cleavage with $K_{trap} \approx 10^4 \text{sec}^{-1}$ [18]), when the time of a solvated electron's occurrence on a nucleotide is not long enough for it to be trapped, the hopping mechanism of transfer is possible. In this case the transfer distance is considerably limited since the probability of the polaron occurrence on the *n*-th site decreases exponentially with increasing *n*.

Experiments in solutions [1] unambiguously demonstrate that the charge can be transferred over 200 and more nucleotide pairs. In explaining this phenomenon, G.B. Schuster and co-authors [1, 19, 20] used a concept of a large-radius polaron, but, as stated above, this concept is in conflict with the picture of fully localized solvated polarons, in the case of a DNA in a solution.

To our knowledge these questions still remain to be answered. This highlights the necessity of developing a theory which would predict the distribution of radical cations along the duplex in the course of their migration over DNA. In experiments reported in refs. [21–25] and refs. [19, 20] on migrations of radical cation (hole) $G^+$ in DNA placed into a solution the charge could be detected by nucleophilic water trapping of $G^+$ which led to strand cleavage products $P_G$, $P_{GG}$, $P_{GGG}$, … at different positions of the radiolabelled strand. The main assumption that we will use to explain the experiments [19–25] is that due to slow reaction rate between $G^+$ and water the radical cations are not trapped by the surrounding medium for sufficiently long time and tunneling or reversible multistep hopping process results, which leads to equilibrium dynamic distribution of the positive charge due to stabilization of the radical cations at different DNA sites. The experiments of interest can be explained if we believe that the quasi-equilibrium state considered is established for the time shorter than that of solvation. For this reason a hole is localized on guanines with the probability determined by quasi-equilibrium distribution of $P_G$, $P_{GG}$, $P_{GGG}$. We will think that in the quasi-equilibrium state the hole occurs in the polaron nonsolvated state, i.e. in equilibrium with the deformation of the DNA duplex that it induces.

To calculate the quasi-equilibrium stage of the distribution of the hole polaron over individual nucleotides at finite temperature we will proceed from the Holstein model defined by the Hamiltonian [26]:

$$\widehat{H} = \sum_{i,j} \nu_{ij} |i\rangle\langle j| - \alpha \sum_i q_i |i\rangle\langle i| + \sum_i k \frac{q_i^2}{2} \qquad (1)$$

where $\nu_{ij}$ are the matrix elements of the hole transition from the $i$-th to the $j$-th site (Fig. 1); $\alpha$ is the coupling constant of the hole interaction with displacement $q_i$ at the $i$-th site, $k$ is the elastic constant. In this model a nucleotide is considered as an individual site and $q_i$ has the meaning of a change in the distance between individual nucleotides in a pair caused by the emergence of a hole at the site.

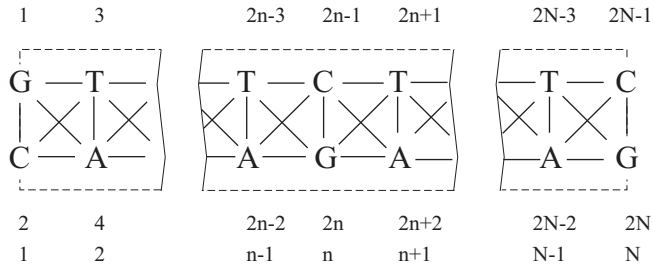

**Fig. 1.** Transitions of a hole between neighboring nucleotides in DNA duplex. Numbers of nucleotides; at the bottom are numbers of nucleotide pairs.

In calculations we used the same values of $\nu_{ij}$ as in paper [15] (Table 1 from [15]). The quantity $\alpha$ is equal to 0.13 eV/E (which is close to the value obtained by quantum-chemical calculations [27]). The quantity $k$ was taken to be equal to 0.062 eV/E$^2$ (which leads to the characteristic frequency of oscillations of a nucleotide pair $\approx 10^{12}$sec$^{-1}$, with the nucleotide pair mass being $\approx 10^{-21}$g).

Knowing a set of steady states, i.e. solutions of Schrödinger equation $\hat{H}|\psi^{(k)}\rangle = E_k|\psi^{(k)}\rangle$, we determine the statistical sum $Z = \sum_k e^{-E_k/T}$ for finite temperature T, and knowing canonical Gibbs ensemble $P_k = e^{-E_k/T}/Z$ we find the population densities of the sites

$$P(j) = \sum_k |\psi_j^{(k)}|^2 e^{-E_k/T} / Z, \quad j=1, 2, \ldots, 2N. \tag{2}$$

For the case of a homogeneous Poly G/ Poly C chain, calculations by formula (2) for T = 300K demonstrate that in the quasi-equilibrium state a polaron can be found at any guanine of the chain with the same probability: $P_{Gn} = 1/N$. If a quasi-equilibrium state does not manage to establish during the time of the hole solvation, the distribution of the strand cleavage products will be non-uniform exponentially decreasing from the site at which the hole occurred at the moment $t = 0$.

Now let us consider the case of regular nucleotide chains. The authors of [20] present distributions of intensities of the products of water–radical cation $G^+$ interaction for oligonucleotides:

$$\begin{pmatrix} GGA\ldots A \\ CCT\ldots T \end{pmatrix}_5 \begin{matrix} GG \\ CC \end{matrix}, \text{ number of A/T pairs } m: 1 \leq m \leq 7, \tag{3}$$

and for oligonucleotides:

$$\begin{pmatrix} GGT\ldots T \\ CCA\ldots A \end{pmatrix}_5 \begin{matrix} GG \\ CC \end{matrix}, \text{ number of T/A pairs } m: 1 \leq m \leq 5. \tag{4}$$

In all the cases calculation by formula (2) for T = 300K yields a quasi-uniform distribution of $P_{GG}$. In the experiment, however, the uniform distribution of $P_{GG}$ is observed in oligonucleotides (3) only for $m \leq 2$. This result is in complete agreement with the model assumed if we believe that for $m > 2$ a quasi-equilibrium state of a hole has no time to establish in oligonucleotides (3). The reason is that the time during which a hole travels a fragment of three A/T pairs that present a wide potential barrier for it, is much longer than the time during which it travels one or two A/T pairs. In the case of bridges of T/A pairs (4) a quasi-equilibrium state has no time to establish for as few pairs as $m > 1$, since the oxidation potential of thymine exceeds that of adenine and a barrier of two T/A pairs in sequence (4) turns out to be higher than that of two A/T pairs in sequence (3).

Now let us consider the case of irregular sequences. In the general case, in irregular sequences a non-uniform distribution of a hole over guanines will be established. This heterogeneity, however, may be caused by two different reasons. In the first case it will be associated with the lack of quasi-equilibrium (kinetic model). In the second case it is attributed to the establishment of a quasistationary state of the hole which will be heterogeneous due to heterogeneous distribution of nucleotides in the duplex. As an example we refer to the results of ref. [22] where the distribution of intensities of the products of water – radical cation $G^+$ interaction is presented for the oligonucleotide:

$$\begin{pmatrix} GTT \\ CAA \end{pmatrix}_4 \begin{matrix} GGG \\ CCC \end{matrix}. \tag{5}$$

Fig. 2 illustrates the distribution of a charge on duplex (5) calculated by formula (2) for T = 300K from the first 204 solutions of Schrödinger equation $\hat{H}|\psi\rangle = E|\psi\rangle$. In Fig. 2 the peaks correspond to guanines $G_j$, $j = 1, 7, 13, 19, 25, 27, 29$ in terms of the numeration of

Fig. 1. In Fig. 2 asterisks mark intensities from ref. [22] normalized on 1. Since the heterogeneous charge distribution calculated from canonical Gibbs ensemble is close to the experimental one we may conclude that the experiment is close to equilibrium.

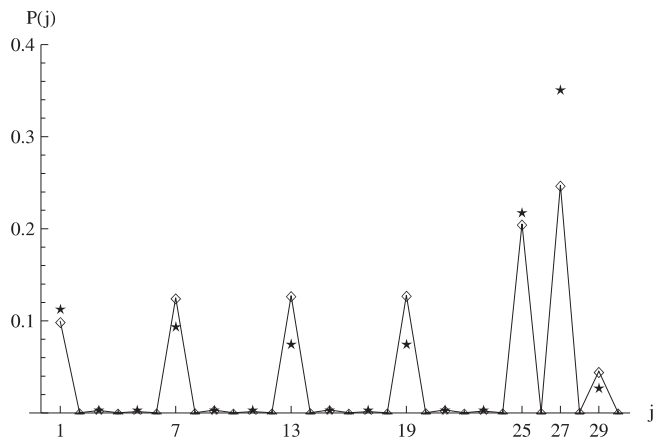

**Fig. 2**. Distribution of hole probabilities $P(j)$ over sites for T = 300K; $j$ is the site number in terms of numeration of Fig. 1. Diamonds – calculation, asterisks – experiment [22].

According to [16, 17], consideration of solvation reduces to inclusion of the solvation energy $S_i$ approximately equal to $\varsigma_i^0(Q_i - 1)$ into diagonal matrix elements $H_{ii}$, where $Q_i$ is the share of the charge density distribution falling on the $i$-th site. The quantity $\varsigma_i^0$ was calculated in [16, 17] to be −1.02eV. Consideration of solvation leads to practically complete localization of a hole and for no parameter values leads to the distribution of Fig. 2.

The results obtained demonstrate that solvation practically does not take part in the formation of a quasi-equilibrium distribution of Fig. 2 which is in complete agreement with the picture of charge transfer in DNA presented above.

This picture is supported by kinetic models. In the kinetic model of ref. [20], the main parameter is a dimensionless quantity $K_{ratio} = K_{hop}/K_{trap}$, where $K_{hop}$ is the rate at which a hole hops between neighboring guanines separated by adenine bases. For a sequence of the form of (3) with $m = 1$ and $m = 2$, the quantity $K_{ratio}$, according to [20], is $K_{ratio}(1) > 200$, $K_{ratio}(2) > 300$. As was found in [20], $K_{ratio} > 200, 300$ simply means that it is too large to be determined by the current method. In the case of sequences (3) and (4), $K_{ratio}$ is estimated by quantum-mechanical calculations to be $K_{hop} \approx 10^{12} \text{sec}^{-1}$ [28]. For $K_{trap} \approx 10^4 \text{sec}^{-1}$ obtained in [18], this yields $K_{ratio} \approx 10^7 \div 10^8$. In the case of such superfast transitions of a hole between bases, the distribution of radical cations among GG steps is determined primarily by its thermodynamic stability on each of the GG steps.

For the sequences with $m > 2$, according to [20], the quantities $K_{hop}$ and $K_{trap}$ turn out to be of the same order of magnitude. This is possible only in the case when the time of solvation (solvent polarization) is less than $K_{hop}^{-1}$. Such a transition takes place in passing on from $m = 2$ to $m = 3$. Quantum-mechanical calculations give an increase in the transition time by about an order of magnitude as the adenine bridge lengthens by one pair [28]. Hence, the solvation time $\tau_s$ is $\approx 10^{-11}$sec. This time is equal in magnitude to the time during which a hydrated electron is formed in water [29]. So, in the case of $m > 2$ we deal with migration of a hole which is already solvated.

Moreover, according to our picture, this is migration of a hole which is strongly localized at nearly one site and is not likely to hop. The quantity $K_{hop}$ falls by nearly eight orders of magnitude. This explains numerous contradictions between numerical modeling of the migration of radical cations in DNA and full-scale experiments. If solvation is not taken into account, numerical modeling leads to the possibility of a very fast and effective charge transfer in DNA which contradicts to many experiments. So, a charge can be transferred in DNA in a solution in two qualitatively different ways. The first way is a superfast charge transition over a large distance, while the second one is a very slow process. The first regime is realized in sequences of a special form in which resonance conditions of tunnel superfast transition take place. It can occur both during the time less than that of a hole solvation and during longer times if such conditions are created as a result of solvation. The second regime, i.e. slow transition is realized after the hole solvation and is made by a hole in a solvated state. In this case, at some stages a superfast transition can also take place if resonance conditions of tunnel transfer are created.

In experiments with dry DNA in homogeneous chains, a large-radius polaron is likely to realize [30]. The hole mobility in this case can be very high [6]. This case is of great interest for creation of DNA-based electronic devices.

The work was supported by RFBR projects № 07-07-00313, № 09-07-12073.